\documentclass{aidb}
\usepackage{graphicx}
\usepackage{balance}
\usepackage[disable]{todonotes}
\usepackage{algorithm}
\usepackage{algorithmicx}
\usepackage{algpseudocode}
\usepackage{subcaption}
\usepackage{balance}   

\usepackage[hidelinks]{hyperref}

\usepackage{amsmath}

\usepackage{placeins}

\vldbTitle{Hands-off Model Integration in Spatial Index Structures}
\vldbAuthors{Ali Hadian, Ankit Kumar, Thomas Heinis}

\begin{document}

\title{Hands-off Model Integration in Spatial Index Structures}

\numberofauthors{3} %

\author{
\alignauthor
Ali Hadian \\
       \affaddr{Imperial College London}\\
       \email{hadian@imperial.ac.uk}
\alignauthor
Ankit Kumar \\
       \affaddr{IIT Delhi}\\
       \email{mt1170727@iitd.ac.in}
\alignauthor 
Thomas Heinis \\
       \affaddr{Imperial College London}\\
       \email{t.heinis@imperial.ac.uk}
}

\maketitle

\begin{abstract}

Spatial indexes are crucial for the analysis of the increasing amounts of spatial data, for example generated through IoT applications. The plethora of indexes that has been developed in recent decades has primarily been optimised for disk. With increasing amounts of memory even on commodity machines, however, moving them to main memory is an option. Doing so opens up the opportunity to use additional optimizations that are only amenable to main memory.

In this paper we thus explore the opportunity to use light-weight machine learning models to accelerate queries on spatial indexes. We do so by exploring the potential of using interpolation and similar techniques on the R-tree, arguably the most broadly used spatial index. As we show in our experimental analysis, the query execution time can be reduced by up to 60\%  while simultaneously shrinking the index's memory footprint by over 90\%. 
\end{abstract}

\maketitle

\section{Introduction}
Spatial data is generated and in need of analysis in many different applications. The proliferation of GPS devices, integrated in mobile devices, along with the growing use of high precision sensors, e.g., LIDAR, to map our surroundings has lead to a deluge of spatial datasets.

Efficiency and speed are key to analyse data in general and spatial data in particular as having timely, actionable insight is crucial. No wonder analytics has moved into main memory where feasible and while spatial datasets are big in size, many of them still fit into main memory, particularly in case of today's machines where main memory is abundantly available.

\todo[inline]{TODO: Advocate hybrid learned/algorithmic indexes and give IF-Btree as an example. Mention that things are more complicated in multidimensional data, and hence we embedded more ideas into R-tree to make it efficient.}
In addition to simply move the data into main memory, recent research also has developed a novel class of indexes that use machine learning methods at their core~\cite{kraska2018case}. More precisely, simple machine learning models are used in traditional indexes to better approximate the specific data distribution of the data set at hand.

Most notably, machine learning models like interpolation have been used in B+-Trees and similar indexes with considerable success in that the models helped to accelerate queries by 50\%~\cite{hadian2019interpolation} - a considerable improvement for data structures, such as the B+-Tree which have been carefully optimised and tuned for decades~\cite{graefe2011modern}. 

In this paper, we consequently study the use of machine learning techniques in spatial indexes. Straightforward reuse of the techniques used to optimise the B+-Tree~\cite{hadian2019interpolation} is unfortunately not possible as multidimensional data with multiple correlated dimensions is more complex. The higher complexity, however, also gives us more degrees of freedom to adapt the index structure.  Our suggested solution accelerates the spatial data structure using a simple predictive model, but at the same time re-arranges the physical layout such that the predictive models are the most effective. As we show, use of interpolation techniques accelerates queries on spatial data in main memory by up to 1.8X (4.2X on multi-threaded execution) for a variety of datasets. 

The remainder of the paper is organised as follows. We first discuss related work in Section\ref{sec:relatedwork} and then motivate the idea of using simple machine learning models in spatial indexes in Sections \ref{sec:motivation} and \ref{sec:implementation}. We then discuss how we implemented the machine learning models in the spatial indexes in Section \ref{sec:ifrtree}. In Section \ref{sec:experiments} we discuss our experimental setup and, more importantly, the experimental results. We finally conclude the paper in Section \ref{sec:conclusion}.

\section{Related Work}
\label{sec:relatedwork}
A plethora of spatial indexes~\cite{multidimensialaccessmethods} has been developed in recent decades. In the following discussion, we focus on relevant work related to learned indexes for spatial data as well as to interpolation used for other, lower-dimensional indexes.

\textbf{Learned index structures.}
In the past few years, tons of research have been done on using machine learning to optimize database systems~\cite{kraska2019sagedb}, most notably learned index structures~\cite{kraska2018case} in which a machine learning model replaces traditional index structures (such as B-tree and hash tables) for locating the physical position of records. Several learned models have been suggested so different indexing problems, such as 
range indexing~\cite{ferragina2020pgm,galakatos2019fiting}, bloom filters~\cite{mitzenmacher2018optimizing,mitzenmacher2018model},
distributed indexing~\cite{li2019scalable}, and handling updates~\cite{ding2019alex,hadian2019considerations}. 

\textbf{Hybrid learned indexes}. Some hybrid methods have been suggested to integrate machine learning into a well-known algorithmic model such as B+tree~\cite{hadian2019interpolation} or run an auxiliary algorithmic index alongside a learned model~\cite{llaveshi2019accelerating}. Such hybrid approaches ensure that while the learned model accelerates the search, the algorithmic index still guarantees the worst-case scenario when  modelling is not effective for the given dataset, and also help in handling updates.

\textbf{Linear models and interpolation.}
Among all learned index models, simple linear models have been arguably the most common~\cite{kipf2020radixspline, galakatos2019fiting, kipf2020radixspline, llaveshi2019accelerating, hadian2019considerations, ding2019alex}. Even for a generic learned index such as the RMI model that supports a variety of linear and non-linear models~\cite{kraska2018case}, it is experimentally shown that the best configuration  found for most real-world datasets is simply a linear spline model~\cite{kipf2019sosd,marcus2020cdfshop}.
Linear interpolation has been recently studied by some works~\cite{van2019efficiently,graefe2006b,hadian2019interpolation}. Despite that an interpolation model is not as accurate as other linear models such as linear regression, it has the interesting feature that it does not need training, making interpolation an effective choice to be embedded inside hybrid indexes~\cite{hadian2019interpolation,hadian2020madex}. 

\textbf{Learned spatial indexing.} 
Recently, some efforts have been done on accelerating spatial and multidimensional indexes with matching learning techniques. Nathan et al. suggested a specific learned multidimensional index that learns from the query distribution how to optimize a grid index structure~\cite{nathan2020learning}. Also, LISA~\cite{li2020lisa} is a learned disk-resident R-tree. The difference between LISA and our work is that LISA focuses on minimizing the IO on disk by transforming the data into a single dimension using a lattice regression model. However, such data transformation is not worthwhile in a main-memory index due to the closer ratios of CPU and memory access overheads. Following a different line of work, some effort has been made to learning and soft functional dependencies between the dimensions in spatial indexes and exploiting them for performance optimisation on spatial indexes~\cite{ghaffari2020leveraging,ding2020tsunami}.

\section{Motivation} 
\label{sec:motivation}
Different approaches have been developed for indexing multidimensional datasets. One common approach is to use tree structures, based on space-oriented partitioning such as KD-Tree, Octree, Quadtree, or based on data-oriented partitioning like the R-tree. Hierarchical trees are easy to manage and update, and are efficient choices for databases. Arguably the most prominent spatial index is the R-Tree~\cite{guttman1984r}, the de-facto spatial index of a modern DBMS and used in IBM Informix, MySQL, PostgreSQL, Oracle, and PostGIS. However, in main memory, hierarchical trees require excessive pointer-chasing to execute queries. While on disk, for which the R-Tree historically has been developed, the time to follow pointers is insignificant (compared to the overhead of retrieving data from disk) but in main memory the time for chasing pointers contributes significantly to the overall query execution time.

New hardware thus calls for new approaches for designing indexing algorithms. First, the new indexes we design need to reduce latency at the cost of bandwidth, e.g., having bigger nodes in search trees addresses the issue of excessive pointer chasing. Second, given that the compute power of modern CPUs has improved considerably (as opposed to the memory bandwidth), the use of the CPU can and should be increased to reduce the need to access data. Third, new indexes should give preference to computations that are more friendly for new hardware, e.g., preference should be given to arithmetic operations over branch-heavy operators which can deteriorate the pipeline. In this regard, machine learning is an effective tool that facilitates designing hardware-efficient indexes that exploit patterns in data to reduce data retrieval.

While there is great potential for applying advanced machine learning techniques powered by SIMD operations, this paper focuses entirely on linear interpolation --- the simplest possible model --- and surprisingly we show significant performance improvement even in this case.

\section{Interpolation Friendly Spatial Indexes}
\label{sec:ifrtree}
\subsection{Spatial Indexing Principles}

A vast number of spatial indexing methods have been developed in the last decades~\cite{multidimensialaccessmethods}. Arguably the most broadly used spatial indexes are the R-Tree, the KD-Tree and the Octree (and its two dimensional equivalent, the Quadtree). We primarily focus on these indexes in our work, but the techniques described can be used to optimise further spatial indexes as well.

All these indexes work based on the same principles: they recursively partitions space and create a hierarchical structure so as to guide query execution. At the bottom of the trees, the data (or pointers to the data) is stored in the leaf nodes. The other nodes, the non-leaf nodes, are used to guide query execution and store the partitioning of the space. 

The major difference between the R-Tree and the KD-tree as well as the Octree is the approach to space partitioning used. Historically developed for use on disk, the R-Tree uses a data-oriented partitioning, i.e., it partitions space such that each leaf node neatly fills a disk page (and therefore optimises the use of the disk bandwidth). Doing so leads to a partitioning entirely driven by the distribution of the data. The partitioning is stored in the non-leaf nodes, i.e., each non-leaf node $N$ stores a number of pairs $<M, P>$. $M$ is the minimum bounding rectangle (MBR) enclosing all MBRs (or spatial objects in case of the leaf node) of one child node $C$ while $P$ is the pointer to $C$.

The other approaches use a space-oriented partitioning, i.e., they partition the space recursively. The KD-tree cycles through the dimensions and at every level takes the median in one of the dimensions and splits space accordingly into two half spaces. It is essentially a binary tree, split on the median of the data indexed, i.e., using a hyperplane to split the space. Instead of only two children, the Octree and the Quadtree use eight and four children respectively and split the space/subtree in the middle rather than on the median.

In case of all indexes, a spatial range or point query (essentially a range query with extent zero), is executed by starting at the root node and traversing down the hierarchy using the MBRs (in case of the R-tree) or hyperplanes (in all other cases) in non-leaf nodes to ultimately find the answer to the query stored in the leaf nodes.

Primarily due to their broad use in practice, we focus on the R-tree, KD-Tree and the Octree (and its two dimensional equivalent, the Quadtree). While the R-Tree has historically been developed to speed up execution of queries on data stored on hard disk, it is still competitive when used in main memory as we will also show experimentally.

\subsection{Approach \& Contribution}
In line with previous work on the B+-tree we therefore aim to use a model to predict the location of the records, primarily on the leaf nodes. Doing so will accelerate scanning the leaf nodes. At the same time, as scanning the leaf nodes becomes substantially faster, the optimal setting for the leaf node size is likely to be bigger as well. Bigger leaf node sizes in turn mean fewer leaf nodes which means fewer pointers and thus less pointer chasing, thus speeding up access further.

One particular issue on making a learning-augmented spatial index is to understand which part of the spatial index can be augmented with prediction models. Unlike one-dimensional range indexes such as B+-trees that have a nearly identical layout on all levels, the R-Tree has entirely different nodes as the leaf and non-leaf (internal) nodes. Internal nodes of the R-Tree store the MBRs of their children along with pointers to the children while the leaf nodes only store data, i.e., points. The MBRs stored in internal nodes consist of four values in 2D (and six in 3D) and the optimisation potential for internal nodes is therefore limited as we, for example, only can interpolate on one dimension/value yet we store four or six dimensions (plus a pointer per child node). We consequently focus on optimising access to leaf nodes where we can, for example, interpolate on one dimension but only have to store one additional dimension (or two in 3D).

\subsubsection{Defining a Storage Order} 
The IF-X indexes enrich their unmodified equivalent indexes with insights from learned indexes. A major challenge, however, is how to define an order for the records so that we can best predict. In a range index like B-tree, records are sorted by the key, hence the learned model simply learns the cumulative density function which maps the value of a key to the position of the key in the sorted array. In multi-dimensional data, however, there is no such total order defined for the records as there are multiple different dimensions over which the data can be sorted. One approach is to learn a projection $\mathcal{L}: \mathcal{R}^d \rightarrow \mathcal{R}$ that maps each d-dimensional record into a single dimension, hence recording the data points~\cite{kraska2019sagedb}. Nonetheless, such a conversion is computationally expensive, and this approach has only been effective for disk-based R-trees where the CPU time is negligible compared to IO cost~\cite{li2020lisa}. 

\begin{figure}
  \begin{minipage}[b]{.49\textwidth}
    \centering
    \includegraphics[width=\linewidth]{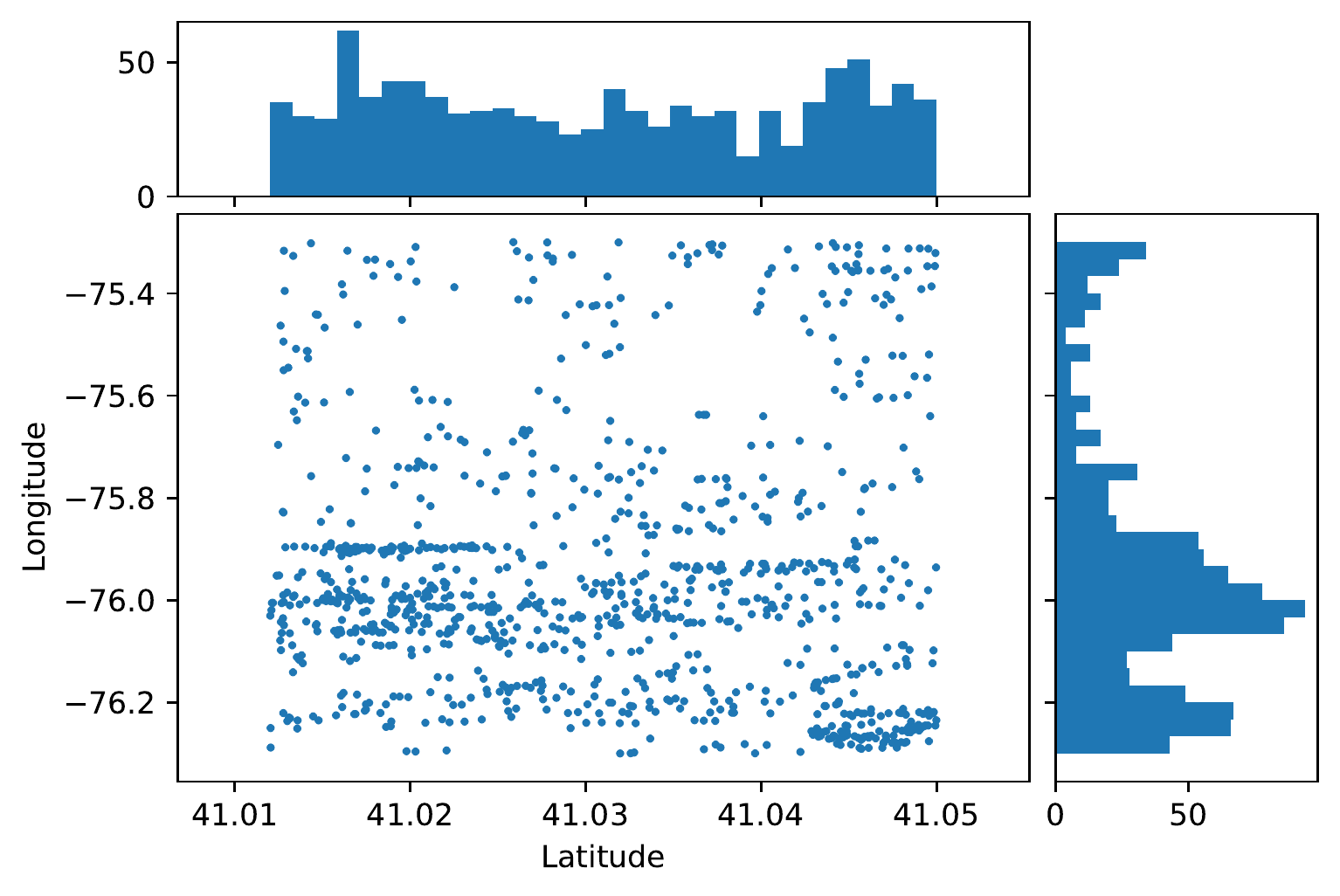}
    \subcaption{Data distribution in a leaf node and the 1D mappings }
    \label{fig:choosing-best-axis-2d}
  \end{minipage}
  \begin{minipage}[b]{.49\textwidth}
    \centering
    \includegraphics[width=\linewidth]{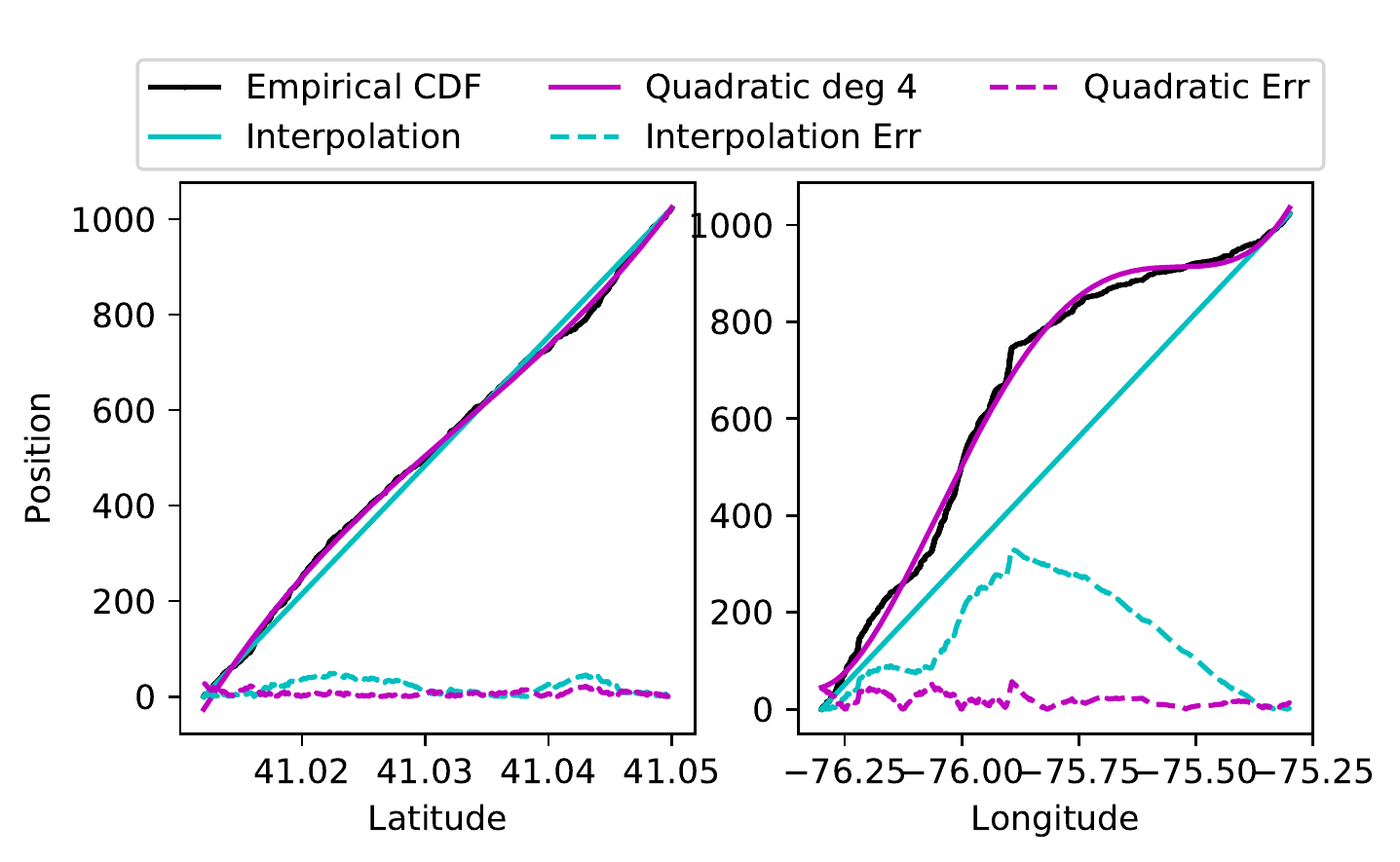}
    \subcaption{Evaluating the prediction error on different dimensions}
    \label{fig:choosing-best-axis-errors}
  \end{minipage}
  \caption{Choosing the most predictable axis in IF-X}
  \label{fig:choosing-best-axis}
\end{figure}

We take a different approach that does not need data transformation: sorting based on one of the existing dimensions. We choose for each leaf node individually the dimension in which it is the most predictable, i.e., the dimension for which the error of the model is the smallest. Figure~\ref{fig:choosing-best-axis-2d} shows the data distribution of a leaf node containing 1024 2D records (spatial points), and the 1D projection of the points in each of the dimensions. 

Figure~\ref{fig:choosing-best-axis-errors} illustrates this by showing the CDF of the values in each dimension, along with two prediction models: linear interpolation (based on minimum and maximum values), and a quadratic polynomial with degree of 4 ($pos = ax^4 + bx^3 + cx^2 + dx+ e$) that is fit to the data. The dashed lines show the error of the models. In the example leaf, the latitude values are more uniform and more predictable for both models.

Note that more complex models such as polynomial or RBF models have a higher capacity to fit the CDF and are able to predict the positions more accurately. However, such complex models need to store more parameters for prediction and are also slower to compute, hence we only use linear interpolation. An additional benefit of interpolation is that there is no need to run an expensive training process.

\subsubsection{Maximum Versus Average Error}. The choice of the best dimension for physical storage order also depends on the local search algorithm. Once the location for a key is predicted by the model, a local search  is performed around the predicted position to find the correct result of the query. Local search in a learned index can be done using binary search, which requires specifying a range, or by other algorithms that do not need a specified range, such as linear search and exponential search. The common practice in learned indexes is to keep track of the \textit{maximum prediction error} per node, say $\Delta$, so that a search can be done on $[\hat{pos}(x) \pm \Delta]$  where the result is guaranteed to be. Therefore, the complexity of the search is in the leaf node is $O(\log\Delta)$. Binary search is very effective and can be implemented branch-free~\cite{dirty2017performance}. However, if $\Delta$ is large, binary search can incur multiple TLB- and cache-misses, which can drastically reduce the performance. One effective solution to counter this issue is to check if the \textit{average prediction error}, say $\bar{\Delta}$, is far less than the maximum prediction error ($\Delta$). In this case, it is more efficient to use linear search or exponential search, which rely on the average error instead of the maximum error. Linear search, for example, starts  from the predicted location and scans towards left or right until it finds the first result belonging to the query, where it \textit{stops} searching. 

In this case, it is more efficient to use linear search, which is $O(\bar{\Delta})$ or exponential search ($O(\log \bar{\Delta}))$. This also affects the choice of the dimension. If the search algorithm is unbounded (linear or exponential search), then the dimension which has the smallest $\bar{\Delta}$ is chosen as the storage order. Therefore, the most predictable dimension (storage order) in each leaf node is chosen with respect to the specific model being used as well as the in-leaf search algorithm.

\section{Implementation}
\label{sec:implementation}

\subsection{Data Structures}

\subsubsection{Leaf Node Layout}
As discussed earlier, IF-X indexes use the same index structure as of their basic/plain indexes, except on the leaf level. None of the indexes considers any specific order for the records stored in each leaf node. Therefore, if the boundaries of a leaf node overlaps with the query, the indexes examine (compare) all records in the leaf node and select the ones that match with the query. As scanning large nodes is generally costly, in-memory the indexes are best configured have smaller leaf nodes which increases the depth of the tree. 

\begin{figure}
    \centering
    \includegraphics[width=\linewidth]{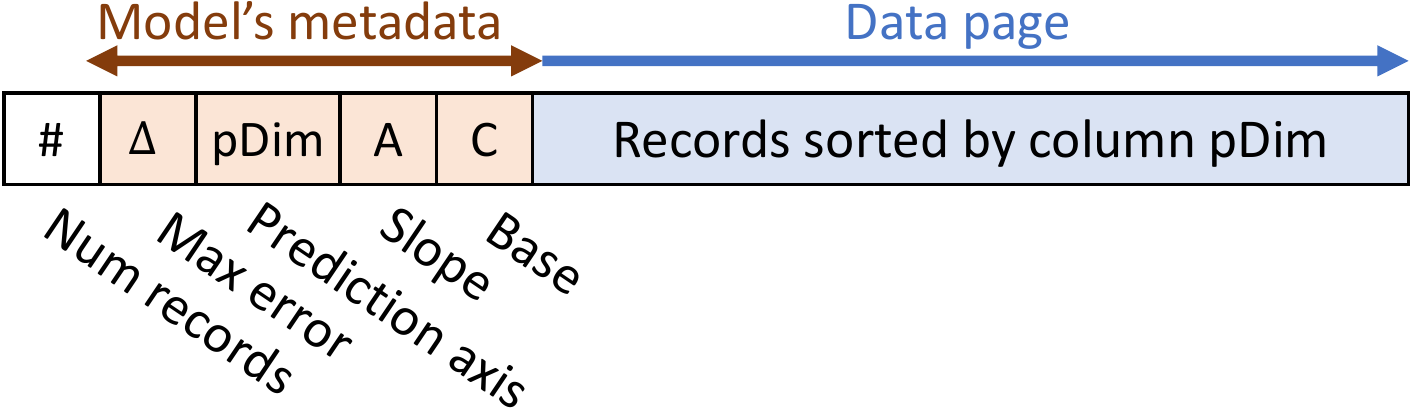}
    \caption{Layout of leaf nodes of IF-X indexes}
    \label{fig:leaf-layout}
\end{figure}

IF-X indexes, on the other hand, sort the records in each leaf node based on the best order using which the interpolation error is minimized, as explained earlier. Our goal is to store all necessary information in the header of the leaf node, such that no extra memory lookup or excessive computation is required other than loading the data pages. Figure~\ref{fig:leaf-layout} shows the layout of the leaf node. The leaf contains the number of records $K$, the most predictable dimension used as the storage order ($\texttt{pDim}$), and model parameters (the slope $A$ and base $C$ of the line for linear interpolation). Also, we store the maximum error in case that binary search is used as the local search algorithm. Nothing else needs to be stored.

Note that in case of linear interpolation, it is not essential to store the model parameters in the leaf node. Linear interpolation does not need training and can be done using the minimum and maximum values of the chosen dimension, which can be fetched from the records page. However, accessing both ends of the records' page incurs further memory lookups and TLB/cache misses. Instead, at build time, we consider the minimum and maximum values in each node, pre-compute the model parameters (slope $A$ and the base $C$), and store those parameters in the header of the leaf node. The model parameters for linear interpolation are:

\begin{equation}
\begin{aligned} 
A &= \frac{maxPos - minPos}{maxVal - minVal} \\
B &=  minPos - A \times minVal
\end{aligned} 
\end{equation}

The position of a query point in a leaf node can be predicted as $\hat{pos} = A \times q[pDim] + B$. In fact, slope pre-computation and re-use is shown to be an effective optimization for iterative interpolation search~\cite{van2019efficiently} and we observed the same benefit when interpolation is used as a model.

\subsubsection{Storage Order Implementation}
Evaluating the predictability of each dimension requires sorting data on each dimension ($O(K \log(K))$ where K is the number of records in page), and then training and evaluating the model on that order. In case that linear interpolation is used, the model requires no training and takes a single scan to evaluate, hence the total complexity is $O(d K \log(K))$.

An additional optimization we use is to define a second sort dimension for records that have the same value for the primary storage dimension. In case that many records have the same value in the primary sort dimension, the second dimension allows for a faster linear search among those records with the same value on the primary sort dimension. We go one step further and sort the records lexicographically based on all columns, i.e.,  sorting the records by $[pDim, 0,1,\cdots,d-1]$. This enables further optimization on the scan operator, e.g., reducing branch misprediction in unrolled loops. 

\subsection{Building \& Updating}
We use the STR~\cite{str} bulkloading strategy to partition the data and then use our leaf node creation algorithm to create the leaf nodes. All other nodes are subsequently created using the STR approach. Any other approach to build an R-Tree or the other trees can be used whilst modifying it to use our leaf node creation approach.

Our suggested modification to spatial indexes does not affect the ability of IF-X indexes to handle updates. Indeed, any update strategy can be used in connection with recently proposed methods to outfit one-dimensional learned indexes for update-heavy workloads~\cite{ding2019alex,hadian2019considerations}, extended to multidimensional learned index structures.

\subsection{Querying}
\subsubsection{Point Queries}
Processing point queries is straightforward, the only difference in implementation with IF-X indexes is on the leaf node search. 
Given a query $q={x_0, x_1, \cdots, x_{d-1}}$, we first consider the predictable storage dimension of the leaf ($pDim$) and try to find the first record where $points[i][pDim] = q[pDim]$. To do so, we use the learned model of the leaf node to predict the location of the record. For a linear model (including linear interpolation), the predicted position is $\hat{pos} = A \times points[pDim] + B$. Then, a local search  is performed around the predicted location. Local search can be done using either binary, linear, or exponential search. If the local search algorithm is binary search, the boundary for search would be $pos[\hat{pos} \pm \Delta][pDim]$. Once the first (left-most) record with $points[i][pDim] = q[pDim]$ is found, all records with equal value on $pDim$ are compared with query on all dimension to examine if they match.

Algorithm~\ref{alg:search-in-leaf} describes the search algorithm for point queries. 

\begin{algorithm}
    \caption{Local search in leaf nodes for point query }
    \label{alg:search-in-leaf}
    \begin{algorithmic}[1] %
        \Procedure{LeafSearch}{q, points, model, pDim, sDim} 
        \State  $\hat{pos}$ =  model.predictLoc(q[pDim])
    \If{q[pDim] $>$ points[$\hat{pos}$][pDim]}
            \State pos = search(from=$\hat{pos}-\Delta$, to = $\hat{pos}$, axis=d)
    \ElsIf{q[pDim] $<$ points[$\hat{pos}$][pDim]}
            \State pos = search(from=$\hat{pos}$, to = $\hat{pos}+\Delta$, axis=d)
    \EndIf
    \If{q[pDim] $=$ points[pos][pDim]}
        \State{// Use the second dimension for guiding search}
         \If{q[sDim] $>$ points[pos][sDim]}
            \State pos = LinearSearch(from=pos, direction=left, axis=sDim)
        \Else
        	\State pos = LinearSearch(from=pos, direction=right, axis=sDim)
        \EndIf
        \While{q[pDim,sDim] = pos[pos][pDim,sDim]}
            \If{query = points[pos]}
                \State \Return data[pos]
                \State pos += 1
            \EndIf
        \EndWhile
    \EndIf
    \State \Return null
    \EndProcedure
    \end{algorithmic}
\end{algorithm}

\subsubsection{Range Queries}
In range queries, the query is defined by a rectangle, i.e., $q = [(l_0,u_0),\cdots,(l_{d-1},u_{d-1})]$. Since the records are sorted by $pDim$, we find the first record where $q[pDim].l \geq \;\;$ $points[i][pDim]$. To find such a record, we do the same range search as in range indexes~\cite{kraska2018case}, which is similar to the point query search (Algorithm~\ref{alg:search-in-leaf}), but replaces the equality check with inequality ($\leq$) in the matching constraint. Once the first record is found, we perform a scan from the found position and compare all records with the query constraints, until we reach the first record that hits the upper limit of the query on $pDim$'s dimension, i.e., $q[pDim].u < points[i][pDim]$, after which no further records will match the query.

\section{Experimental Analysis}
\label{sec:experiments}
\subsection{Experimental Setup}
\label{subsec:experimental_setup}
\textbf{Hardware.} Single-core experiments are performed on a system with 16 GB of memory and Intel Core i7-6700 (Skylake), which has four cores and is running at 3.4 GHz with 32 KB L1, 256 KB L2, and 8 MB L3 caches. Multithreaded experiments (Figure~\ref{fig:point_speedup_parallel}) are conducted on an AMD EPYC 7401 server running at 2GHz with Virtualization, having 24 available cores with 32 KB L1, 512 KB L2, and 64 MB L3 caches. Both machines have 16GB of memory.

\textbf{Software.} All indexes are implemented in C++ and compiled with GCC 9.2. Experiments are run on Ubuntu 18.04 with kernel version 4.15.0-65. 

\begin{figure}
  \begin{minipage}[b]{.49\textwidth}
    \centering
    \includegraphics[width=\linewidth]{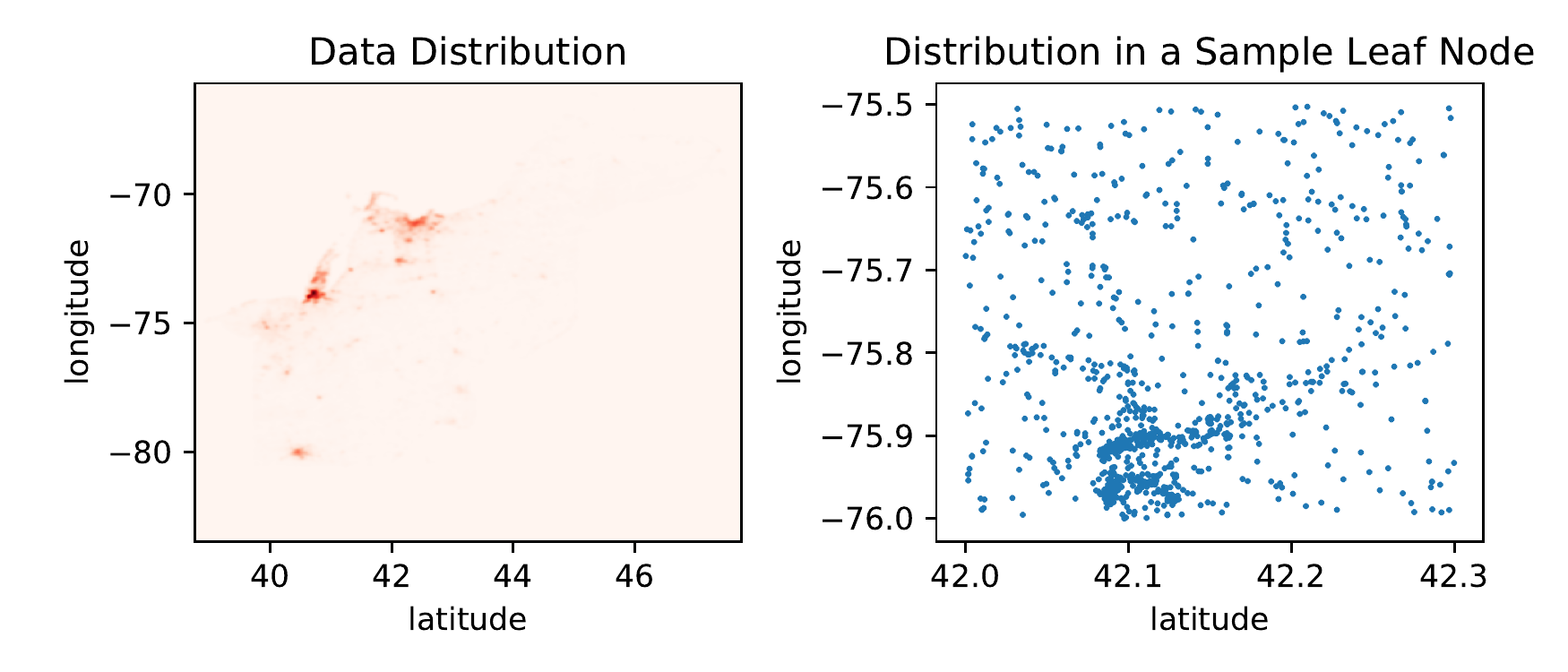}
    \subcaption{OSM (OpenStreetMap, Northeast section)}
    \label{fig:distribution_osm}
  \end{minipage}
  \begin{minipage}[b]{.49\textwidth}
    \centering
    \includegraphics[width=\linewidth]{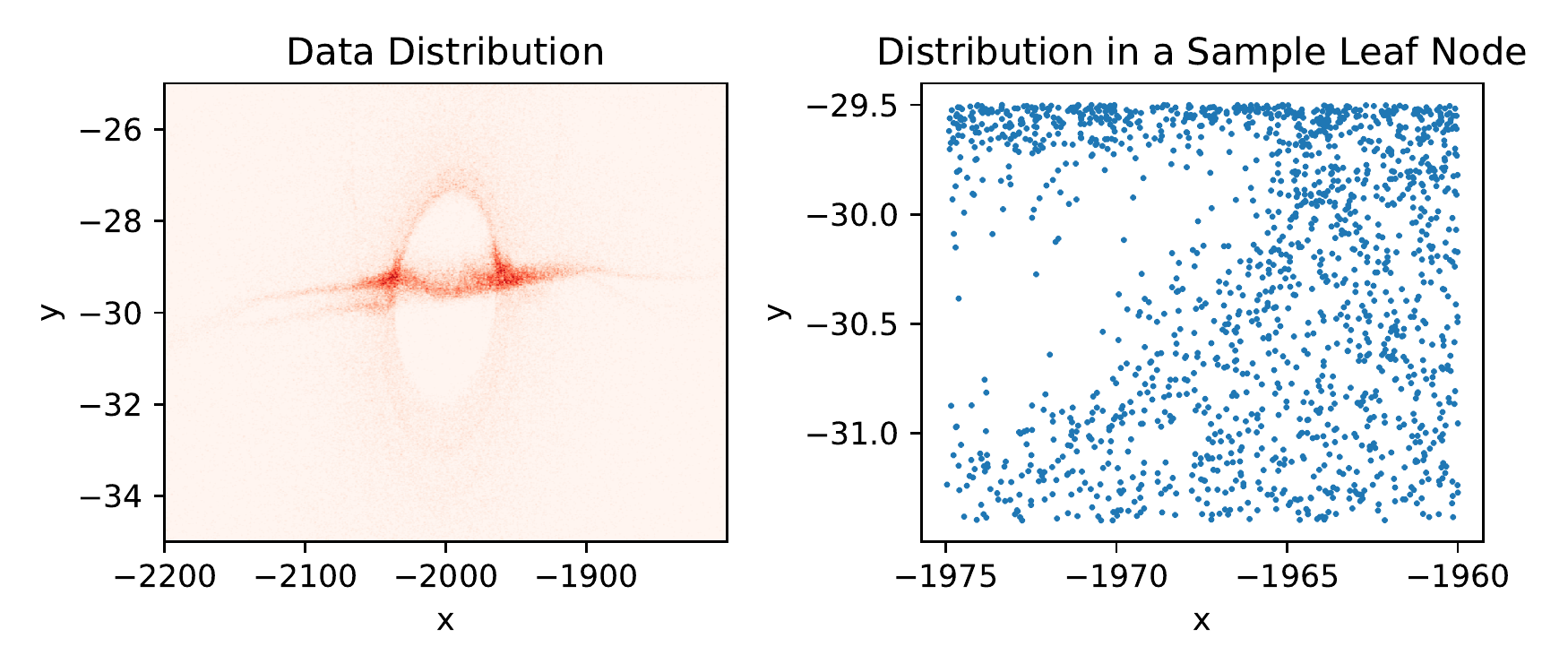}
    \subcaption{3DScans}
    \label{fig:distribution_3dscans}
  \end{minipage}
  \caption{Distribution of the datasets}
  \label{fig:distributions}
\end{figure}

\textbf{Datasets.} 
We used two spatial datasets: \textit{osm}: latitude and longitude of landmarks from the US northeast section of  OpenStreetMap data ~\cite{openstreetmap}, with the timestamp being used as the third dimension in the 3D experiments; and  \textit{3DScans}: A point clouds collected for robotic experiments, built by a 3D scan (containing X,Y,Z coordinates) taken in front of the cathedral in Zagreb, Croatia~\cite{3dscans}. We used $1M$, $10M$ and $100M$ sample records from each dataset. Figure~\ref{fig:distributions} shows the heatmap of data distribution in both OSM and 3DScans data, along with an illustration of data distribution in sample leaf nodes within each dataset. We used 2D data with 10M records as the default configuration for the experiments. Values from all dimensions are stored as single-precision floating point values.

\subsection{Tuning}
\begin{figure*}
  \begin{minipage}[b]{.5\textwidth}
    \centering
    \includegraphics[width=\linewidth]{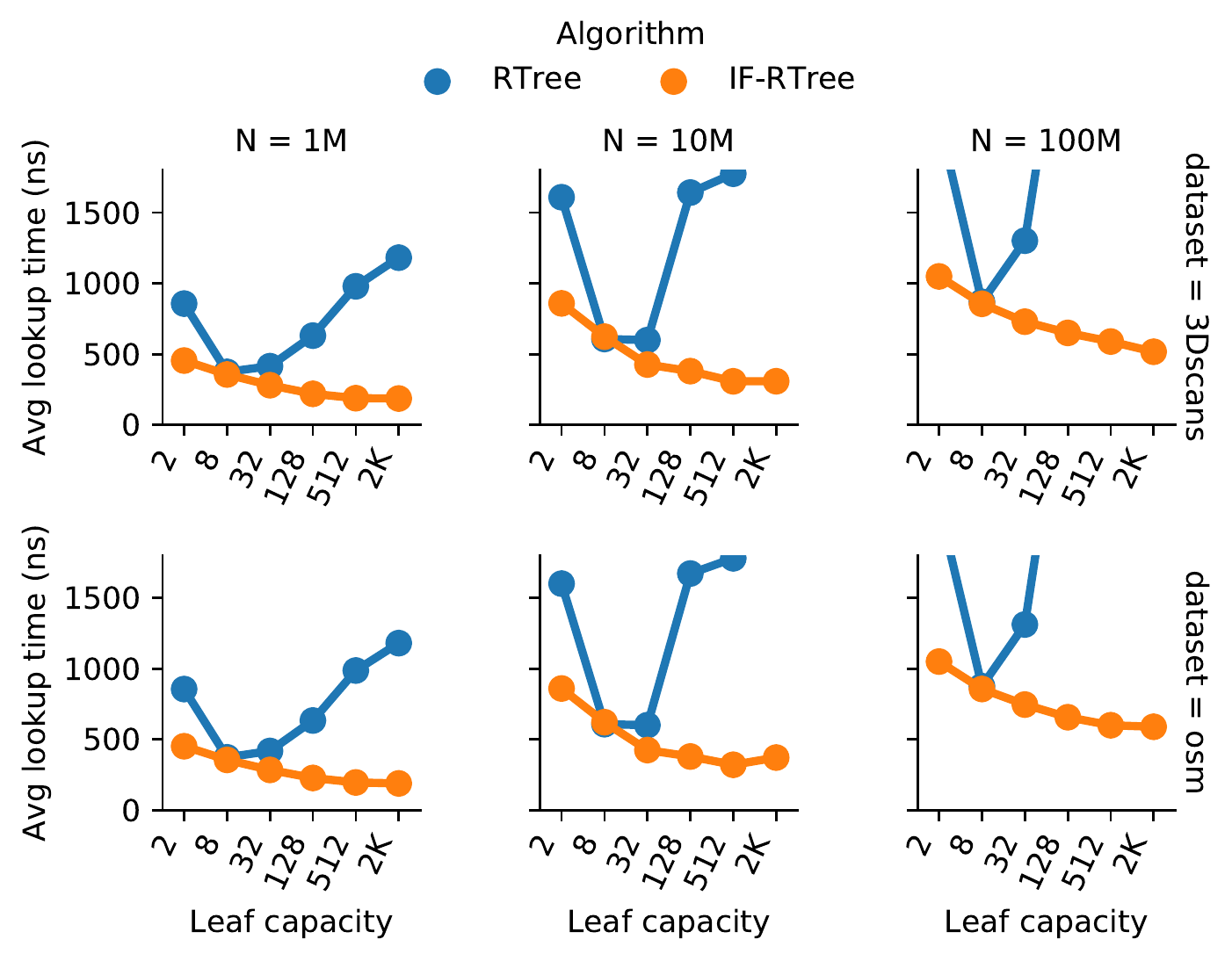}
    \subcaption{2D data}
    \label{fig:point_varmaxcap_2d}
  \end{minipage}
  \begin{minipage}[b]{.5\textwidth}
    \centering
    \includegraphics[width=\linewidth]{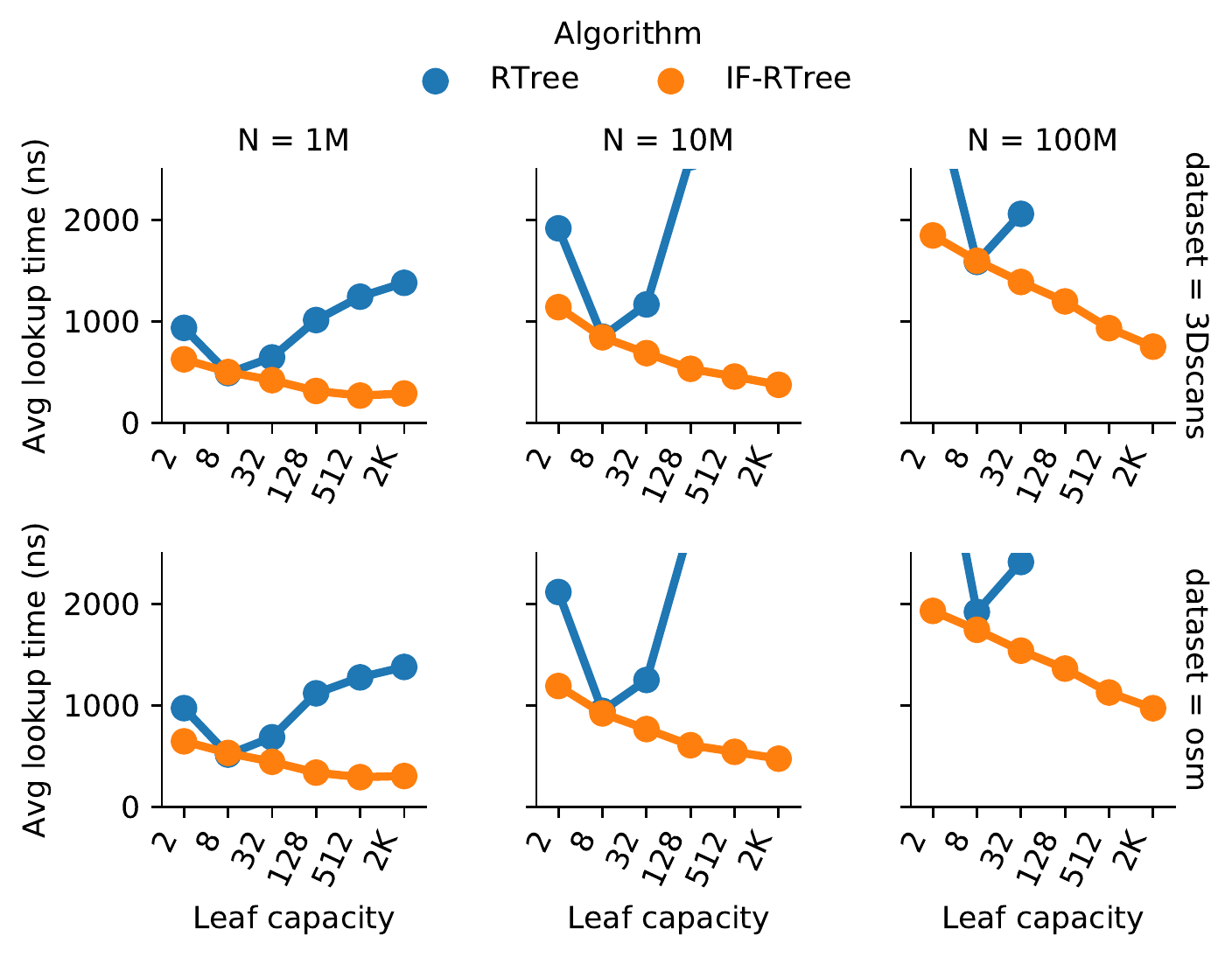}
    \subcaption{3D data}
    \label{fig:point_varmaxcap_3d}
  \end{minipage}
  \caption{Index parameter tuning: effect of leaf capacity on R-tree and IF-RTree indexes}
  \label{fig:point_varmaxcap}
\end{figure*}

Each spatial index is tuned with its internal parameters, most notably with node sizes between 2 and 32K records per node ($2^i, 1 \leq i \leq 15$). Figure~\ref{fig:point_varmaxcap}, shows how the optimization has been done on R-tree index, optimized for point queries. For each data size and query type, we took the optimum leaf size for each index and compare the indexes with best configuration. However, the optimal configuration for each indexing algorithm does not vary much for different dimensions. 

In our default configuration (2D index, 10M records, tuning for point queries), the best leaf sizes found were: R-tree:~16, KD-tree:~128, Octree:~256, IF-RTree:~512, \{IF-KDTree, IF-QuadTree, IF-Octree\}:~2048. The optimal configurations found for 2D and 3D data are similar.

\subsection{Point Queries}

\begin{figure}
\centering
   \begin{minipage}[b]{.49\textwidth}
    \centering
    \includegraphics[width=\linewidth]{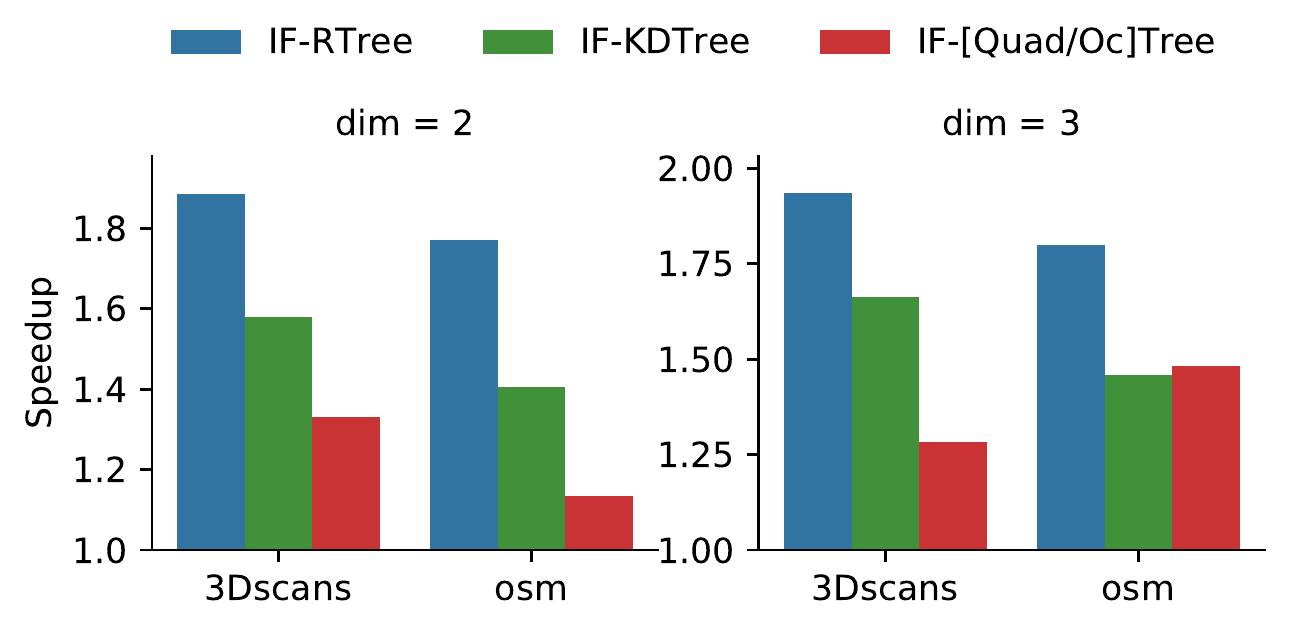}    
    \subcaption{Single-threaded}
    \label{fig:point_speedup_serial}
  \end{minipage}
   \begin{minipage}[b]{.49\textwidth}
    \centering
    \includegraphics[width=\linewidth]{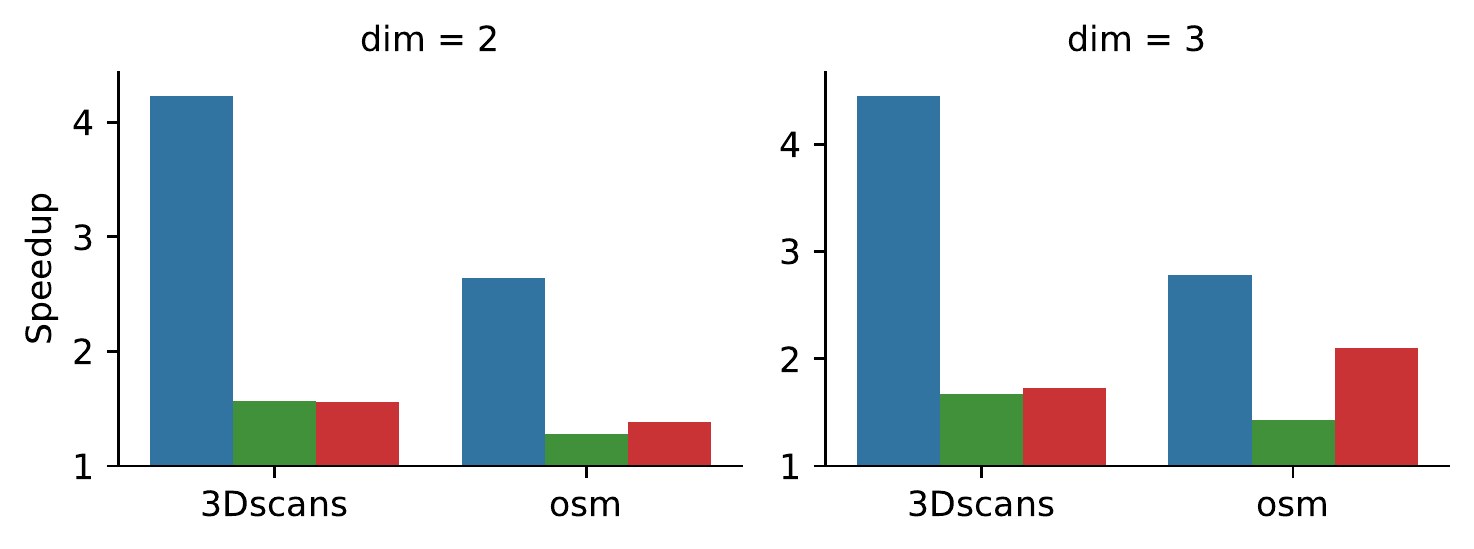}    
    \subcaption{Multi-threaded (24 Cores)}
    \label{fig:point_speedup_parallel}
  \end{minipage}
  \caption{Speedup of the IF-X indexes to their original version for point queries}
  \label{fig:speedup_point_query}
\end{figure}

\textbf{Overall performance}
We first benchmark how well our IF-X indexes can optimize their base versions. The point queries are randomly sampled from the datasets. Figure~\ref{fig:point_speedup_serial} shows the speedup obtained by each IF-X index over its own baseline for each dataset. IF-RTree has the highest speedup (1.8X speedup over R-tree in osm, 1.9X in 3DScans; for both 2D and 3D data). Other algorithms, IF-KDTree and IF-[Quad/Oc]-Tree have speedup between 1.15X and 1.6X for different datasets and dimensions. 

\textbf{Dimensionality}
An interesting observation is that IF-X indexes almost retain their speedup for the 3D data, with even a rise in the speedup of IF-Octree (1.5X speedup on 3D osm, compared to 1.15 in 2D). This is due to the fact that even though the interpolation is done on one dimension, with higher dimensions the IF-X indexes have more freedom to choose the best storage order (separately for each leaf), therefore some of the leaf nodes that are skew in 2D become more predictable when sorted on the third dimensions. We can think of it as, for example, how in Figure~\ref{fig:distribution_3dscans}(right) the x axis makes the data more predictable compared to a one-dimensional index over y only. 

\textbf{Scalability.}
Along with the single-threaded experiments, we compared the speedup of the IF-X indexes in a multi-threaded setting using a 24-core machine (explained in section~\ref{subsec:experimental_setup}). In this setting, we use the total execution of the query batch, hence the measurements and the reported speedup shows the improvement in throughput, not the average query times of individual queries. All the three spatial indexes have benefited from model-assisted acceleration in all datasets, with over 4X speedup in IF-RTree.

\textbf{Performance breakdown.}
We are interested to see where does the speedup of IF-X indexes over the baselines come from. To further analyze the speedups gained for the default configuration (Figure~\ref{fig:point_speedup_serial}-left), we measure the index size and average lookup times.

\begin{figure*}[!htpb]
\begin{minipage}[b]{.3\textwidth}
    \centering
    \includegraphics[width=\linewidth]{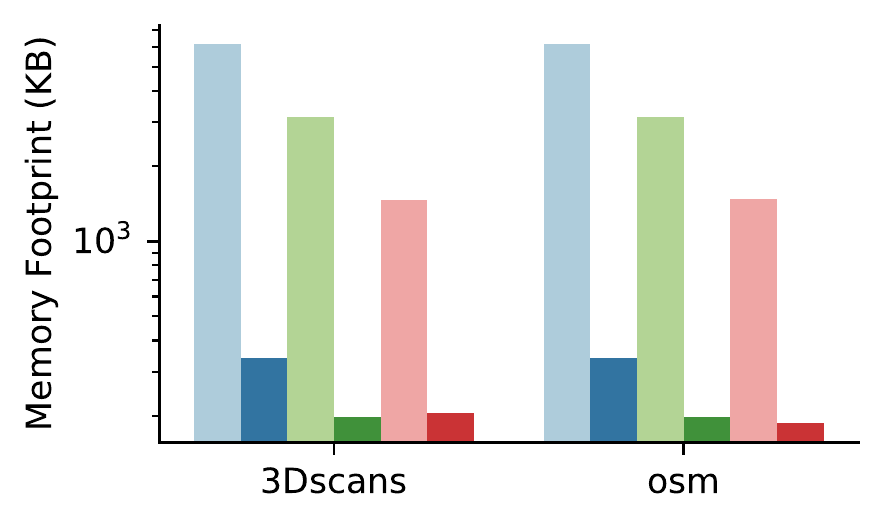}
    \subcaption{Memory footprint}
    \label{fig:bestConf_memory}
  \end{minipage}
  \begin{minipage}[b]{.3\textwidth}
    \centering
    \includegraphics[width=\linewidth]{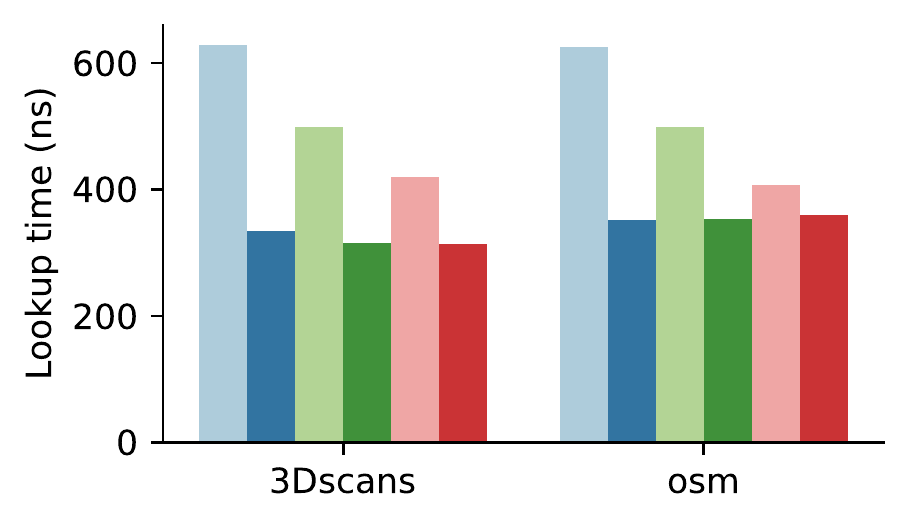}
    \subcaption{Average lookup time}
    \label{fig:bestConf_avg_time_ns}
  \end{minipage}
   \begin{minipage}[b]{.4\textwidth}
    \centering
    \includegraphics[width=\linewidth]{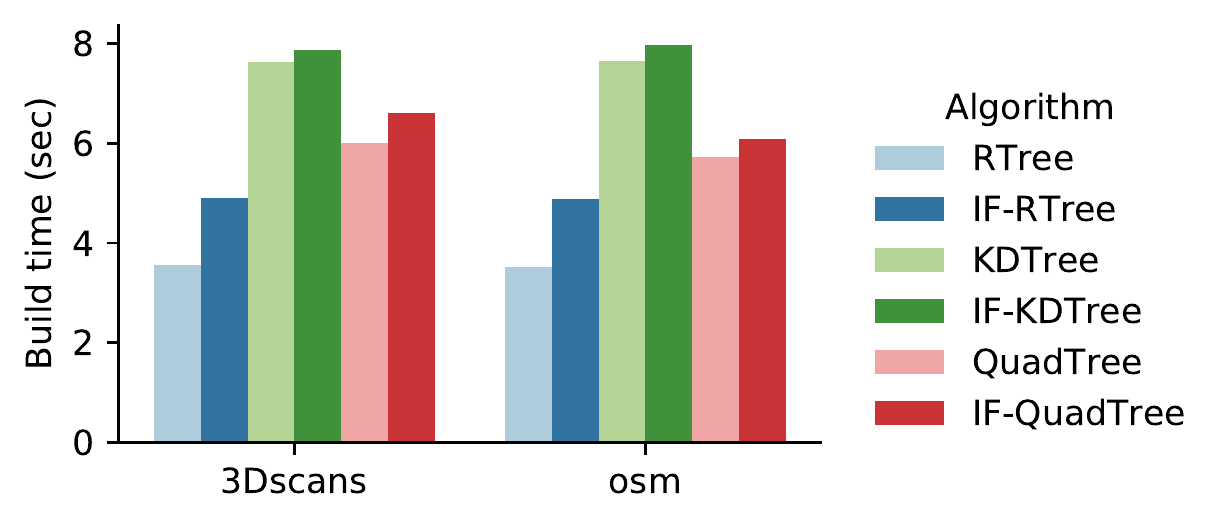}
    \subcaption{Build time}
    \label{fig:bestConf_build_time}
  \end{minipage}
  \caption{Comparison of indexes on their best configuration}
  \label{fig:point_bestConf}
\end{figure*}

\textbf{Memory footprint.} The optimal parameters found in the tuning phase suggest that IF-X indexes prefer much larger leaf nodes than the original X indexes. Larger leaf sizes manifest in smaller footprints in the IF-X index. By footprint, we mean the internal nodes of the indexes and put aside the size of the leaf nodes (which equals the data size).
Figure~\ref{fig:bestConf_memory} shows, in logarithmic scale, the memory usage of each index in its optimal setting for 2D data with 10M records. The difference in memory footprint is significant. For example, IF-RTree takes up only 5.5\% of the size of R-tree, and the other IF-X indexes are no larger than 10\% of their base version. Such a saving in memory footprint is typically enough to push the index up in the memory hierarchy and fit the internal nodes into processor's caches.

\textbf{Average lookup times.} Comparing the lookup times of the algorithms provides insights into why some IF-X indexes have a comparably smaller speedup compared to the others. Lookup times are shown in Figure~\ref{fig:bestConf_avg_time_ns}, which show that all IF-X algorithms derived from different base indexes result in almost identical lookup time. This suggests that the IF-X family from different base methods converge to a similar: A cache-resident index structure along with large, flattened leaf nodes that are sorted to optimize the predictability of the record locations.

 \begin{figure}[H]
  \centering
    \includegraphics[width=\linewidth]{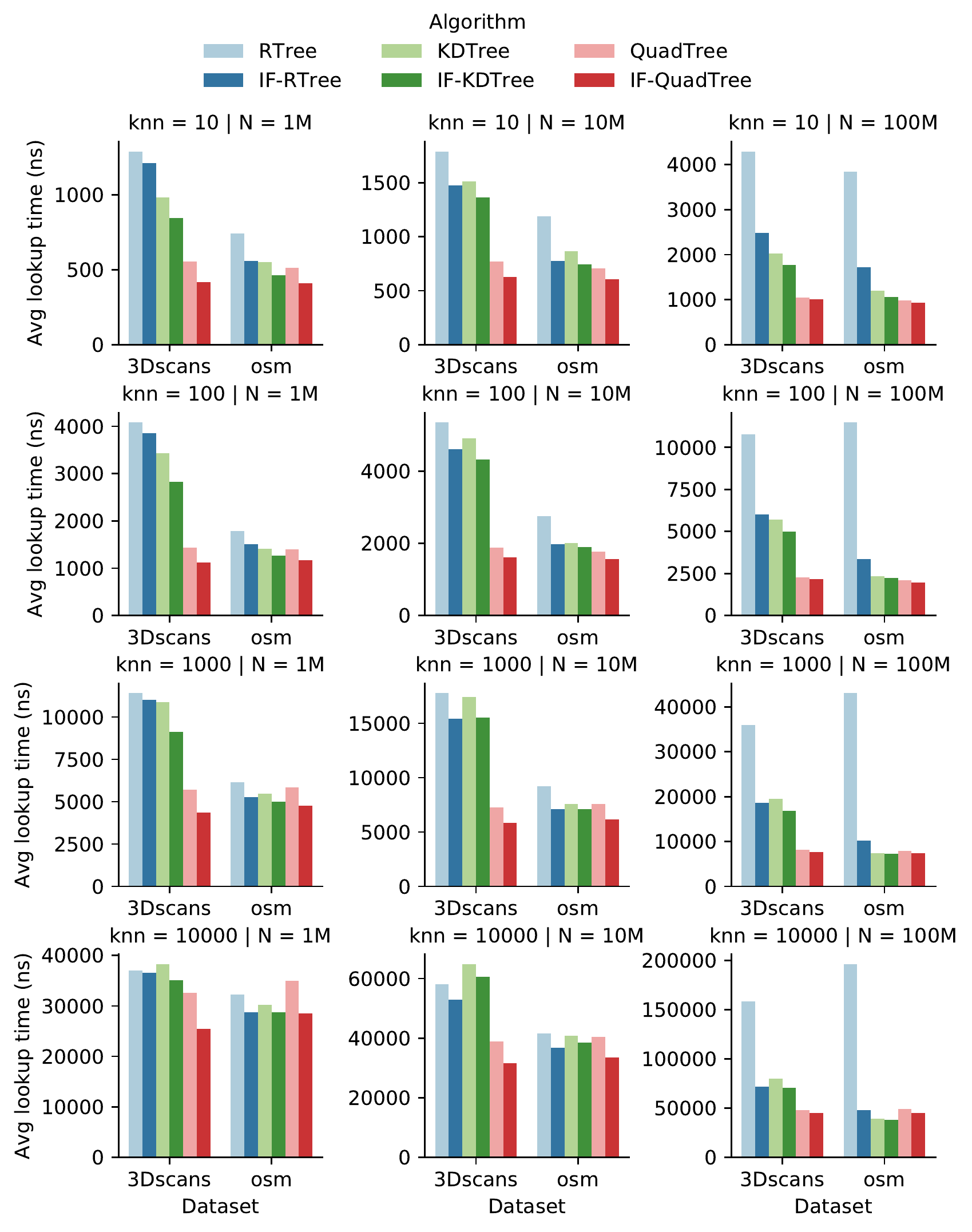}
  \caption{Performance comparison on range queries}
  \label{fig:exp_range_queries_2d}
\end{figure}

\textbf{Build times}
We also report the build times of the indexes. Building IF-RTree takes 37\% longer time than RTree, and the other IF-X indexes take less than 10\% longer to build. Note that each algorithm in the chart has a different leaf size. Unlike the base indexes in which a significant share of the build time is taken on building the internal nodes, the build time of IF-X indexes are mostly taken by sorting the large leaf nodes on multiple axes in order to choose the most predictable storage order. This shows a considerable potential in scaling IF-X methods with parallel execution, because sorting large leaf nodes for storage selection in IF-X can be done independently for each node using multiple threads; while paralleizing the baseline algorithms requires concurrent access control mechanisms on tree indexes and is harder to scale. However, scaling up the build algorithm is not considered in this work.

\balance
\subsection{Range Queries}

For range queries, we considered four sets of queries with different selectivities, $\sigma \in S = \{10, 100, 1K, 10K\}$. We build a KD-Tree for the chosen datasets (for appropriate sizes). For each $\sigma \in S$, we then choose random points in the KD-Tree, and find the set of nearest $\sigma$ points (with euclidean metric), $P$. The MBR of $P$ then gives us the desired query.

Figure~\ref{fig:exp_range_queries_2d} shows the performance of the indexes for range queries. For lower selectivities (small K), query time is mainly affected by the initial lookup of the corner point of the query rectangle and hence the speedups are more similar to that of point queries. For larger selectitivies, however, query times are be mostly taken by the scan operations. Yet, since all IF-X indexes favour larger leaf nodes, the scan operations enjoys a considerable speedup.

\section{Conclusion and Future Work}
\label{sec:conclusion}
In this paper we have integrated learned models into the most broadly used spatial indexes. We dub the resulting indexes IF-X indexes for all of them as they all are based on space-partitioning trees for the organisation of data. Integrating learned models is not straightforward in spatial indexes. Multiple dimensions and the absence of a total order make it challenging to find computationally efficient models that accelerate query execution. As our experiments show, however, our IF-X indexes are consistently considerably faster than the their plain, unmodified equivalents and they also consistently have a smaller footprint. 

In our current work we did not make any assumptions about the query workload. In case that the query workload is known a priori, the leaf creation algorithm can be modified to use the dimension that is most predictable and most helpful for executing the workload. For example, if most queries do not put a constraint on a specific dimension, then that column is not selected for ordering the records. We leave this as future work.

\clearpage

 \balance

\bibliographystyle{abbrv}
\bibliography{bibliography}

\end{document}